\documentclass{PoS}

\usepackage{subfig}
\usepackage[font={small}]{caption}

\title{Detailed studies of hadronic showers and comparison to GEANT4 simulations with data from highly granular calorimeters}

\ShortTitle{Detailed studies of hadronic showers and comparison to GEANT4 simulations}

\author{\speaker{Naomi van der Kolk}\\
	for the CALICE Collaboration\thanks{CALICE Collaboration web page: http://twiki.cern.ch/CALICE}\\
       	Max-Planck-Institute for Physics, Munich, Germany\\
       	E-mail: \email{naomi.van.der.kolk@cern.ch}}

\abstract{
	The highly granular calorimeter prototypes of the CALICE collaboration have provided large data samples with precise three-dimensional information on hadronic showers with steel and tungsten absorbers and silicon, scintillator and gas detector readout. 
	From these data sets, detailed measurements of the spatial structure, including longitudinal and lateral shower profiles and of the shower substructure and time structure are extracted.
	Recent analyses have extended these studies to different particle species in calorimeters with scintillator readout and steel and tungsten absorbers, to energies below 10 GeV in a silicon tungsten calorimeter and have provided first studies of the shower substructure with gaseous readout and unprecedented granularity of $1\times1$~cm$^{2}$ over a full cubic meter. 
	These results are confronted with {\sc{Geant4}} simulations with different hadronic physics models.
	They present new challenges to the simulation codes and provide the possibility to validate and improve the simulation of hadronic interactions in high-energy physics detectors.}

\FullConference{The European Physical Society Conference on High Energy Physics\\
		 22-29 July 2015\\
		 Vienna, Austria}

\begin{document}

\section{Introduction}
\vspace{-2mm}
The CALICE collaboration develops, builds and tests highly granular calorimeter prototypes to evaluate the performance of such detectors for future physics experiments.
The collaboration has built several prototypes for electromagnetic and hadronic calorimeter concepts, using different designs, steel or tungsten absorbers and silicon, scintillator or gas detector readout.
These prototypes and their performance are discussed in a separate contribution to this conference~\cite{PoS_Vladik}.
Currently the following large scale (\textasciitilde1\,m$^{3}$), high granularity ($1\times1$\,cm$^{2}$ -- $3\times3$\,cm$^{2}$) prototypes exist; a scintillator-steel electromagnetic calorimeter (ScECAL), a silicon-tungsten electromagnetic calorimeter (Si-W ECAL), scintillator-steel and scintillator-tungsten analogue hadronic calorimeters (Fe-AHCAL and W-AHCAL), a digital hadronic calorimeter (DHCAL) and semi-digital hadronic calorimeter (SDHCAL) with steel or tungsten absorber and gaseous readout, and a scintillator-steel tail catcher and muon tracker (TCMT).

CALICE prototypes also facilitate the improvement of the understanding of hadronic showers.
A good understanding of hadronic showers is needed for the development and optimisation of Particle Flow Algorithms~\cite{2009_Thomson, 2012_Marshall}.
These algorithms exploit the high granularity of the calorimeters to reach a very good jet energy resolution needed for precision physics~\cite{2001_Brient, 2013_TDR4, 2012_CLIC} .
Hadronic showers have a complex structure and are theoretically not as well understood as electromagnetic showers.
Because of the high granularity of CALICE prototypes, they offer an unprecedented detailed view of hadronic shower structures, which can be used to improve modelling in {\sc Geant4}~\cite{2003_Geant4}.

CALICE prototypes have been operated in hadron test beams at CERN, DESY and Fermilab, which has resulted in large data samples with precise three-dimensional information on hadronic showers.
The collaboration has evaluated the shower energy, shower shape, substructure and time structure of hadronic showers for different particles over a large energy range.
In this conference contribution selected results from recent CALICE studies are shown.
\vspace{-2mm}

\section{Simulation of hadronic showers}
\vspace{-2mm}
Within {\sc Geant4} several theory-driven and phenomenological hadronic interaction models are available~\cite{2010_Apostolakis}.
These models are combined into so-called {\it physics lists}, where they are applied to specific energy ranges.
At low energy ( $\textless10$\,GeV) there are e.g. the Bertini Cascade model (BERT) and the binary cascade model (BIC).
For higher energy there is e.g. the Fritiof String Model (FTFP) and the Quark Gluon String Model (QGSP).
Smooth transitions between models are achieved over a certain energy range by randomly choosing one of the models on an event-by-event basis with a probability that varies linearly with the energy in the transition interval.
The name of the physics list indicates which models are combined in it.
For the last 5 years the most successful physics lists have been QGSP\_BERT and FTFP\_BERT.
The latter is currently the recommended physics list for hadronic shower simulations for the LHC~\cite{2011_Dotti}.
\vspace{-2mm}

\section{Comparison of hadronic showers in simulations and data}
\begin{figure}
  {\centering
    \subfloat[\label{subfig:interactionfraction}]{\includegraphics[height=0.19\textheight]{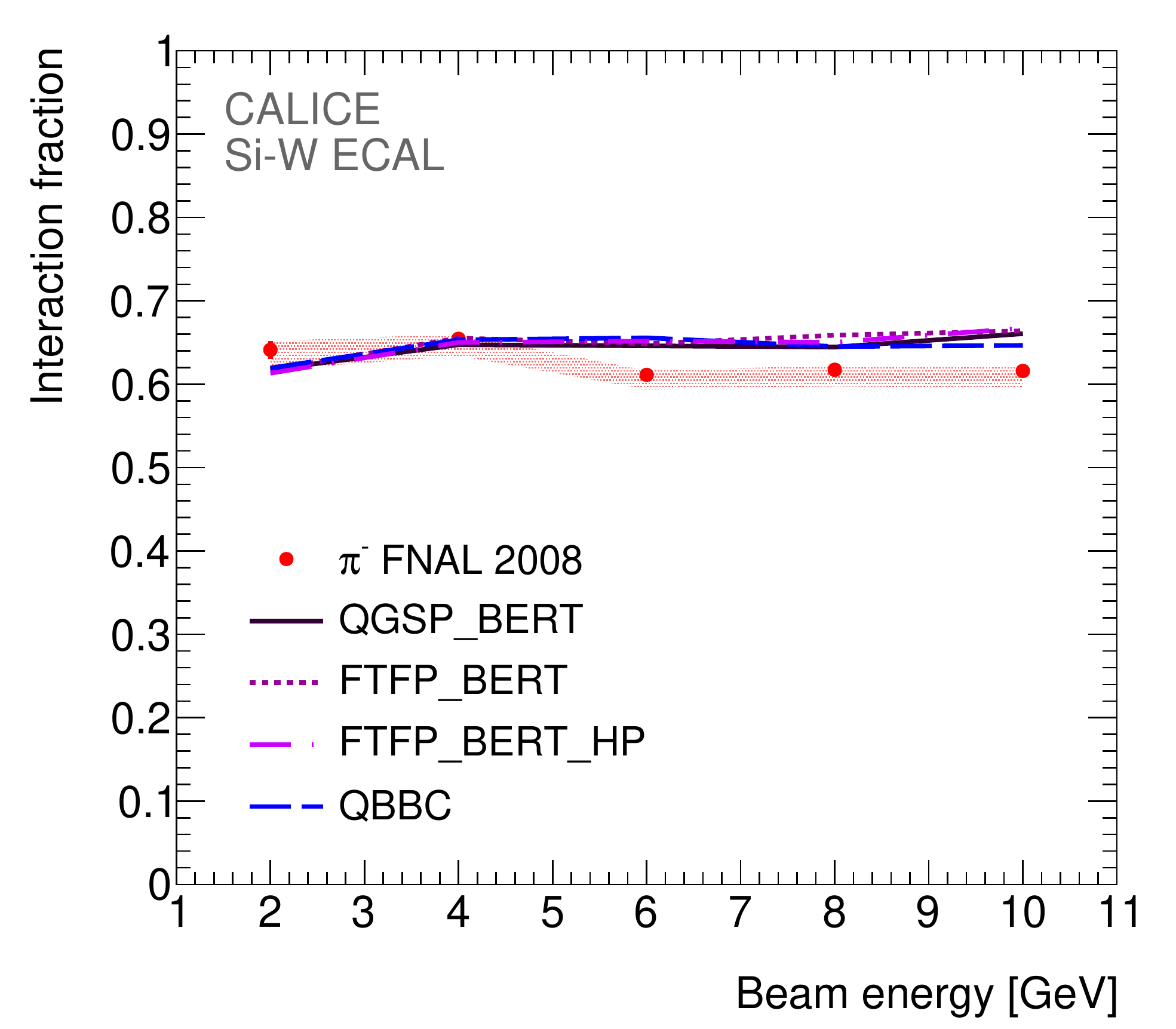}}
    \subfloat[\label{subfig:pioninteractionlength}]{\includegraphics[height=0.19\textheight]{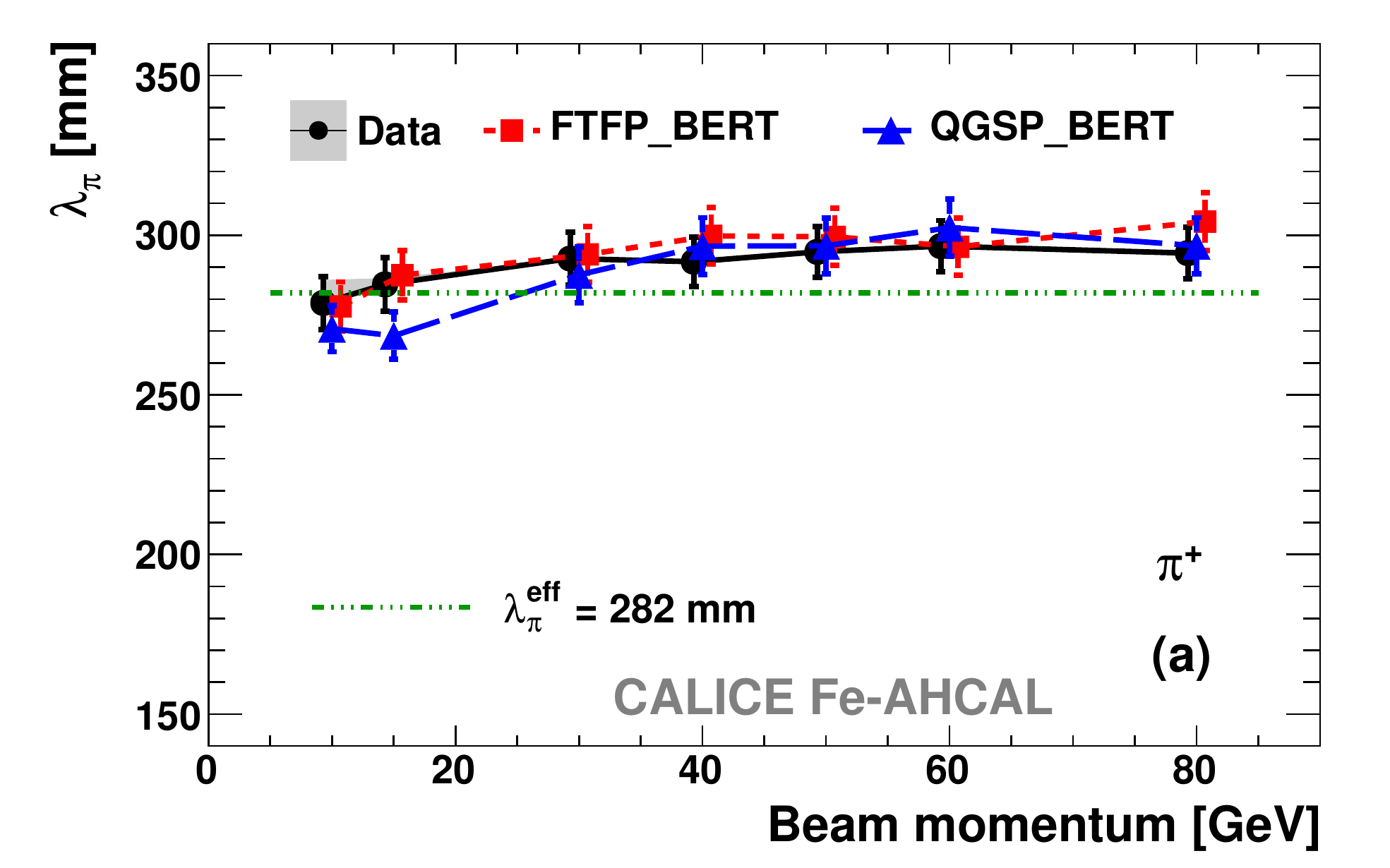}}
    \caption{(a) Interaction fraction for pions between 2\,GeV and 10\,GeV measured in the Si-W ECAL~\cite{2015_SiWECAL}. (b) Pion interaction length measured in the Fe-AHCAL for energies between 10\,GeV and 80\,GeV~\cite{2015_FeAHCAL}. In both figures several physics lists ({\sc Geant4}.9.6p01) are compared to the data.}
  }
\end{figure}

\vspace{-2mm}
In the Si-W ECAL prototype, which has a total thickness of approximately 1 interaction length, the fraction of pions that have interacted and started an hadronic shower has been identified based on an algorithm to find the position of the first hadronic interaction~\cite{2015_SiWECAL}. 
This fraction is shown in Fig.~\ref{subfig:interactionfraction} together with the predictions of several physics lists which are close to the data, but overestimate in the range from 6\,GeV to 10\,GeV.

From the distribution of the position of the first hadronic interaction, the interaction length of the incoming particle can be reconstructed.
It has been determined for pions and protons in the Fe-AHCAL~\cite{2015_FeAHCAL}.
Figure~\ref{subfig:pioninteractionlength} shows the pion interaction length. 
The agreement of the simulations with the data is at the 5\% level.
\vspace{-2mm}
\vspace{-2mm}

The energy deposited in the calorimeter by an incoming hadron is consistent with linearity for energies between 10\,GeV and 100\,GeV for pions and protons in e.g. the Fe-AHCAL~\cite{2013_FeAHCAL} and the W-AHCAL~\cite{CAN044, 2014_WAHCAL} . 
The Monte Carlo simulation describes the data to better than 10\%.

\begin{figure}
  {\centering
    \subfloat[\label{subfig:radialenergy}]{\includegraphics[height=0.21\textheight]{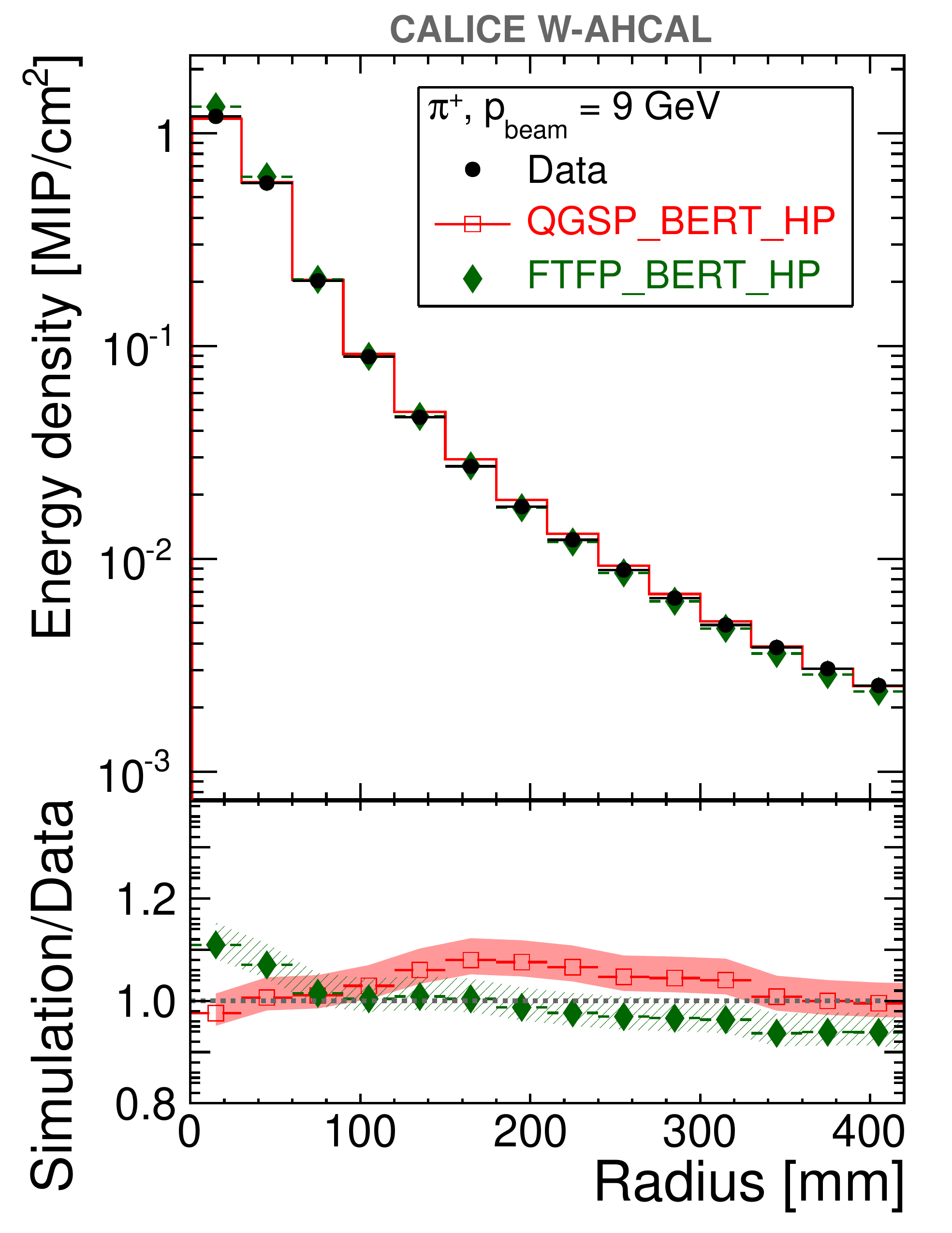}}
    \subfloat[\label{subfig:meanshowerradius}]{\includegraphics[height=0.19\textheight]{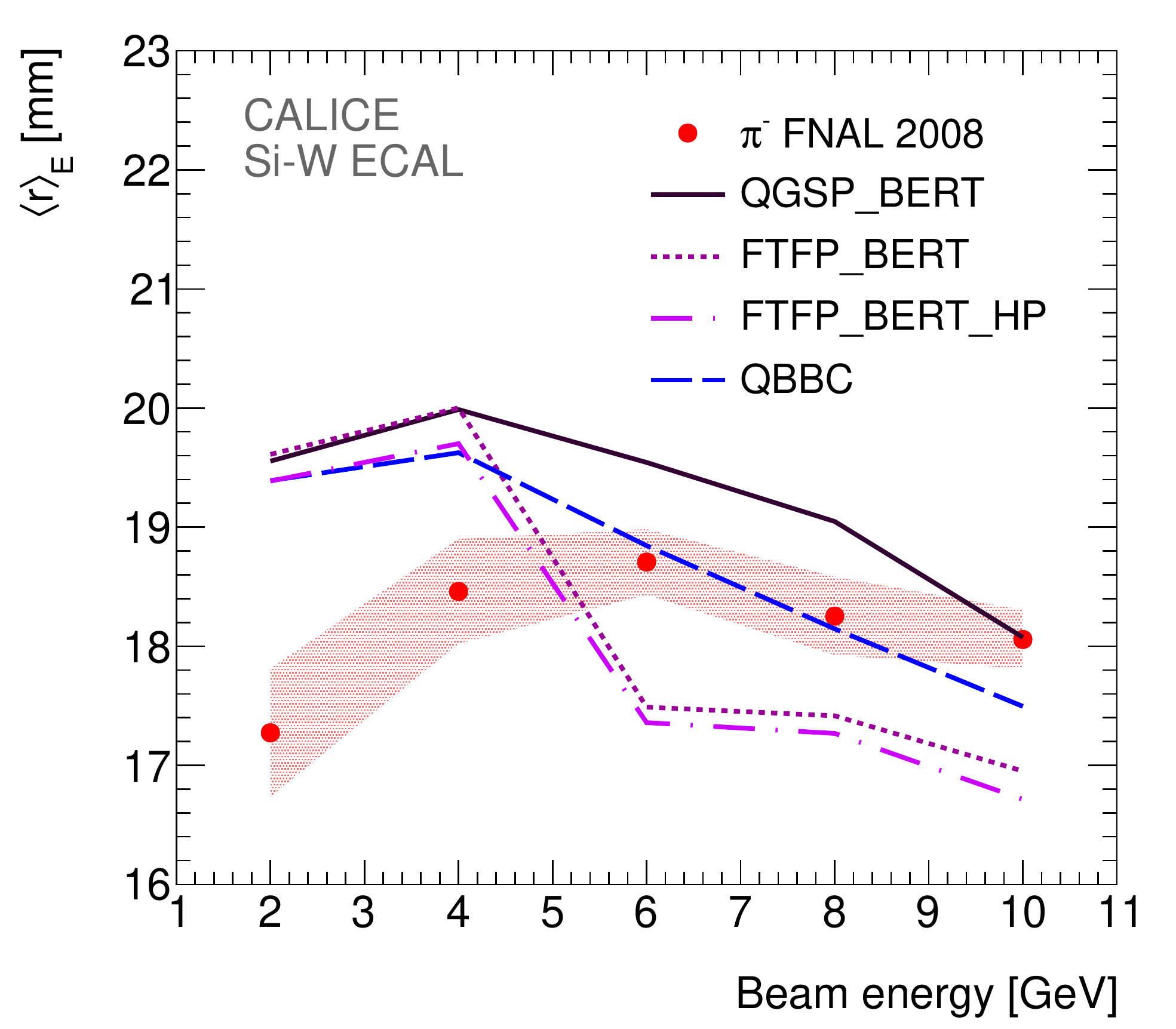}}
    \subfloat[\label{subfig:meanradiusproton}]{\includegraphics[height=0.19\textheight]{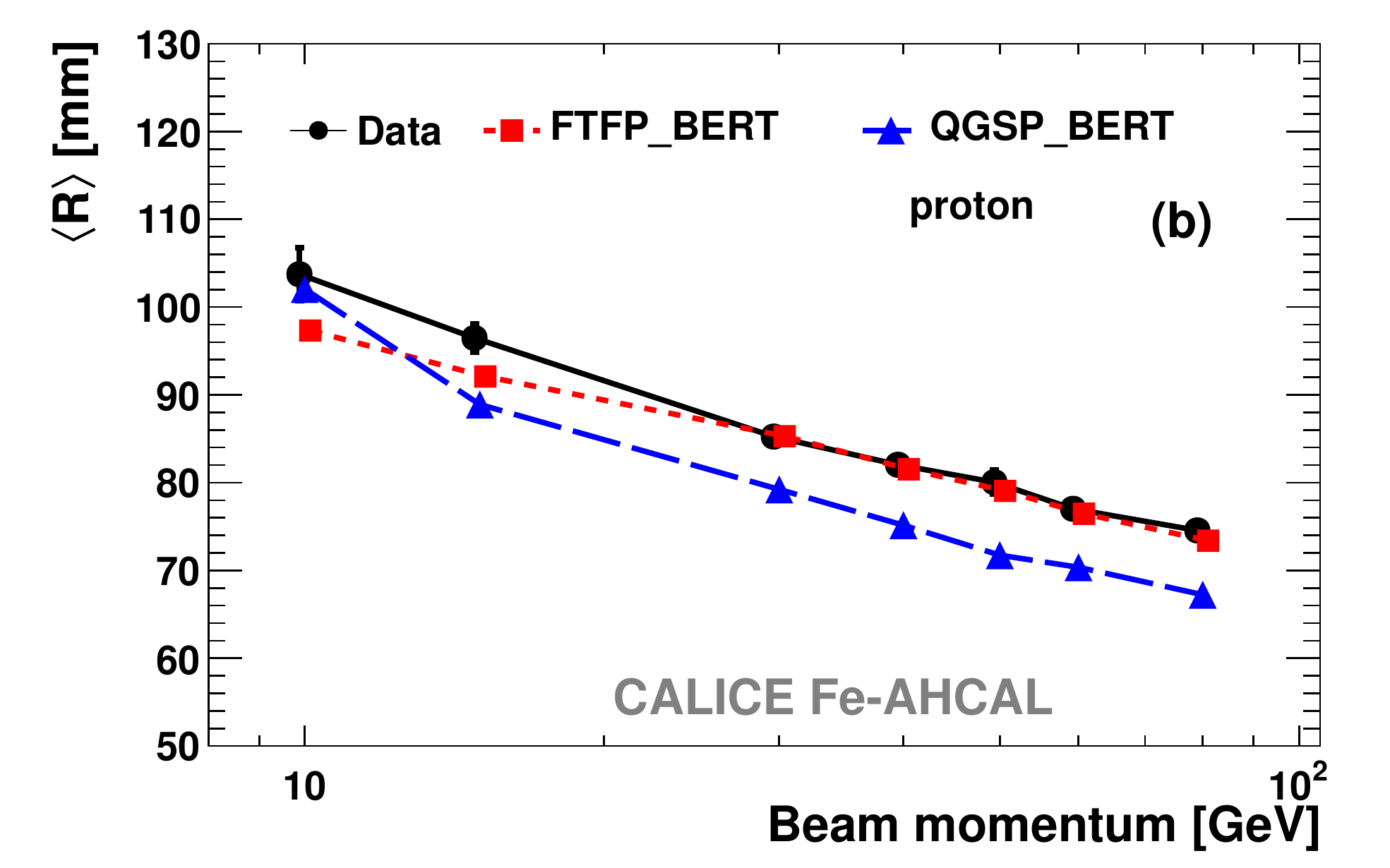}}
    \caption{(a) Radial energy profile for pions at 9\,GeV measured in the W-AHCAL~\cite{2014_WAHCAL}. (b) Mean shower radius for pions measured in the Si-W ECAL for energies between 2\,GeV and 10\,GeV~\cite{2015_SiWECAL}. (c) Mean shower radius for proton showers in the Fe-AHCAL for energies between 10\,GeV and 80\,GeV~\cite{2015_FeAHCAL}. Several physics lists ((a) {\sc Geant4}.9.6p02, (b)(c) {\sc Geant4}.9.6p01) are compared to the data.}
  }
\end{figure}

The width of hadronic showers is evaluated by the radial energy distribution, the energy density as a function of the radial distance to the shower centre of gravity, shown in Fig.~\ref{subfig:radialenergy} for pions at 9 GeV measured in the W-AHCAL~\cite{2014_WAHCAL}.
From these kind of profiles the mean shower radius is extracted.
At low energies simulations in the Si-W ECAL have shown that this observable is sensitive to the model transitions in the physics lists~\cite{2015_SiWECAL}. There is clear discontinuity between the measurement points at 4\,GeV and 6\,GeV in the FTFP\_BERT physics list (the transition from BERT to FTFP occurs between 4\,GeV and 5\,GeV), and a milder one between the 8\,GeV and 10\,GeV points for QGSP\_BERT (the transition between BERT and LEP occurs between 9.5\,GeV and 9.9\,GeV) as can be seen in Fig.~\ref{subfig:meanshowerradius}.

At higher energies (above 10\,GeV) the mean radius decreases with energy, because of the increase of the electromagnetic component of the hadronic shower, which is more compact.
Similar behaviour is seen for pions and protons~\cite{2014_WAHCAL, 2015_FeAHCAL}, although proton induced showers tend to be wider and longer than pion induced showers, due to their smaller electromagnetic shower component.
Figure~\ref{subfig:meanradiusproton} shows the mean shower radius for protons in the energy range of 10 to 80 GeV.
The high level of agreement, generally within 20\%, between simulations and data is due to many recent improvements in the models, to which CALICE data has also contributed.
Still, improvements are possible as the simulated showers are about 10\% denser than real showers.
The best physics lists to describe this observable are FTFP\_BERT and FTFP\_BERT\_HP.

\begin{figure}
  {\centering
    \subfloat[\label{subfig:longitudinalprofileECAL}]{\includegraphics[height=0.21\textheight]{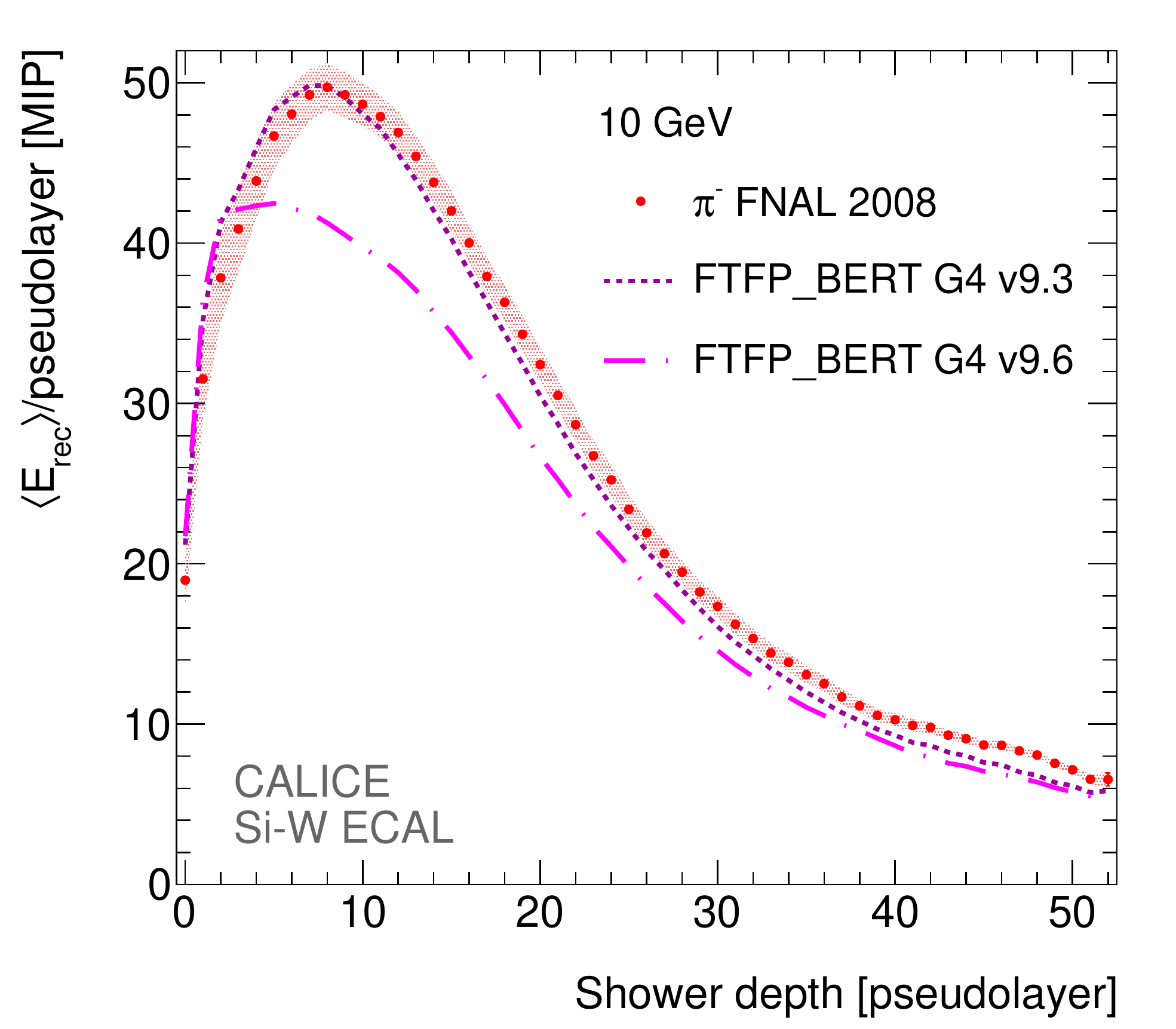}}
    \subfloat[\label{subfig:longitudinalprofileHCALproton}]{\includegraphics[height=0.21\textheight]{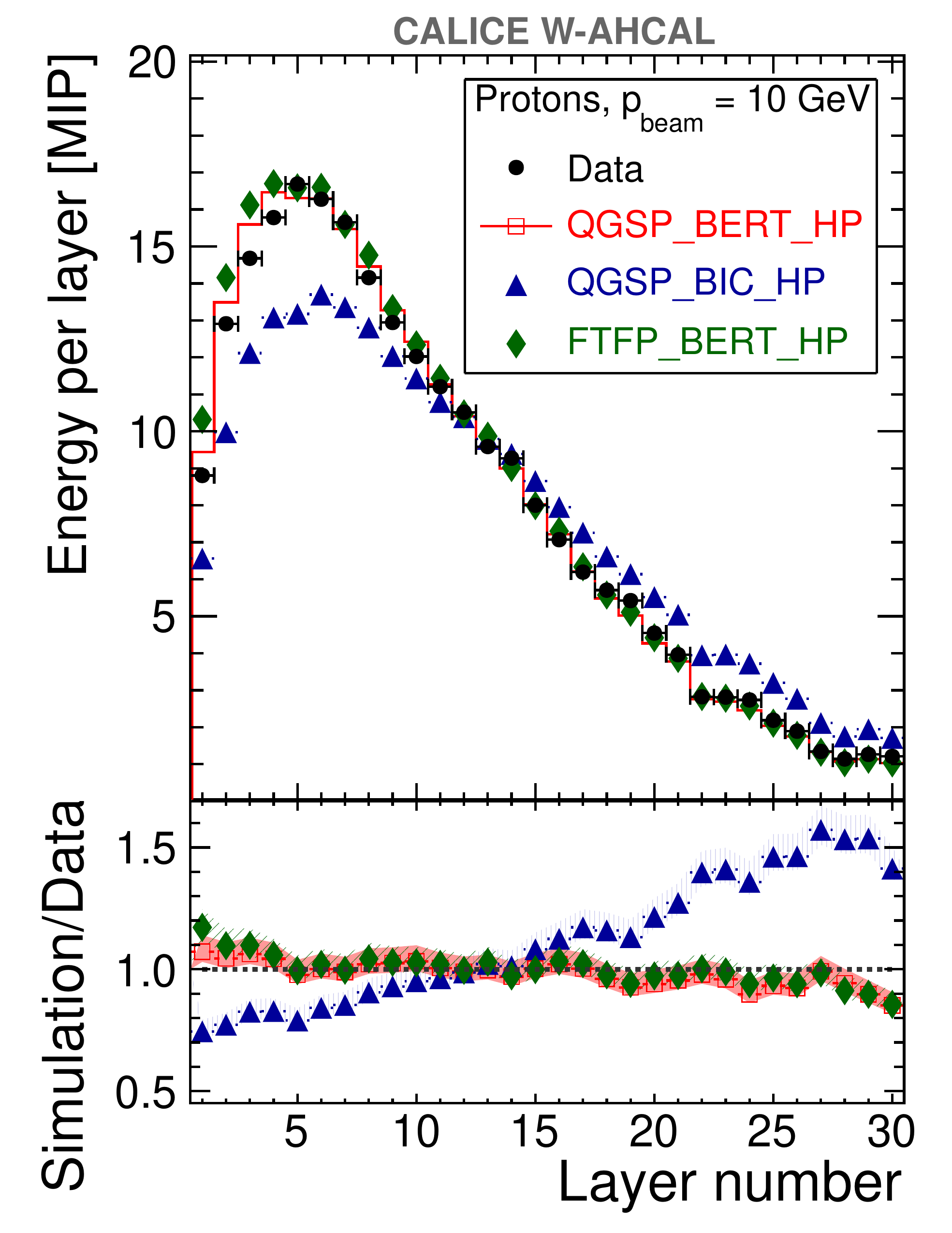}}
    \subfloat[\label{subfig:longitudinalprofileHCALpion}]{\includegraphics[height=0.21\textheight]{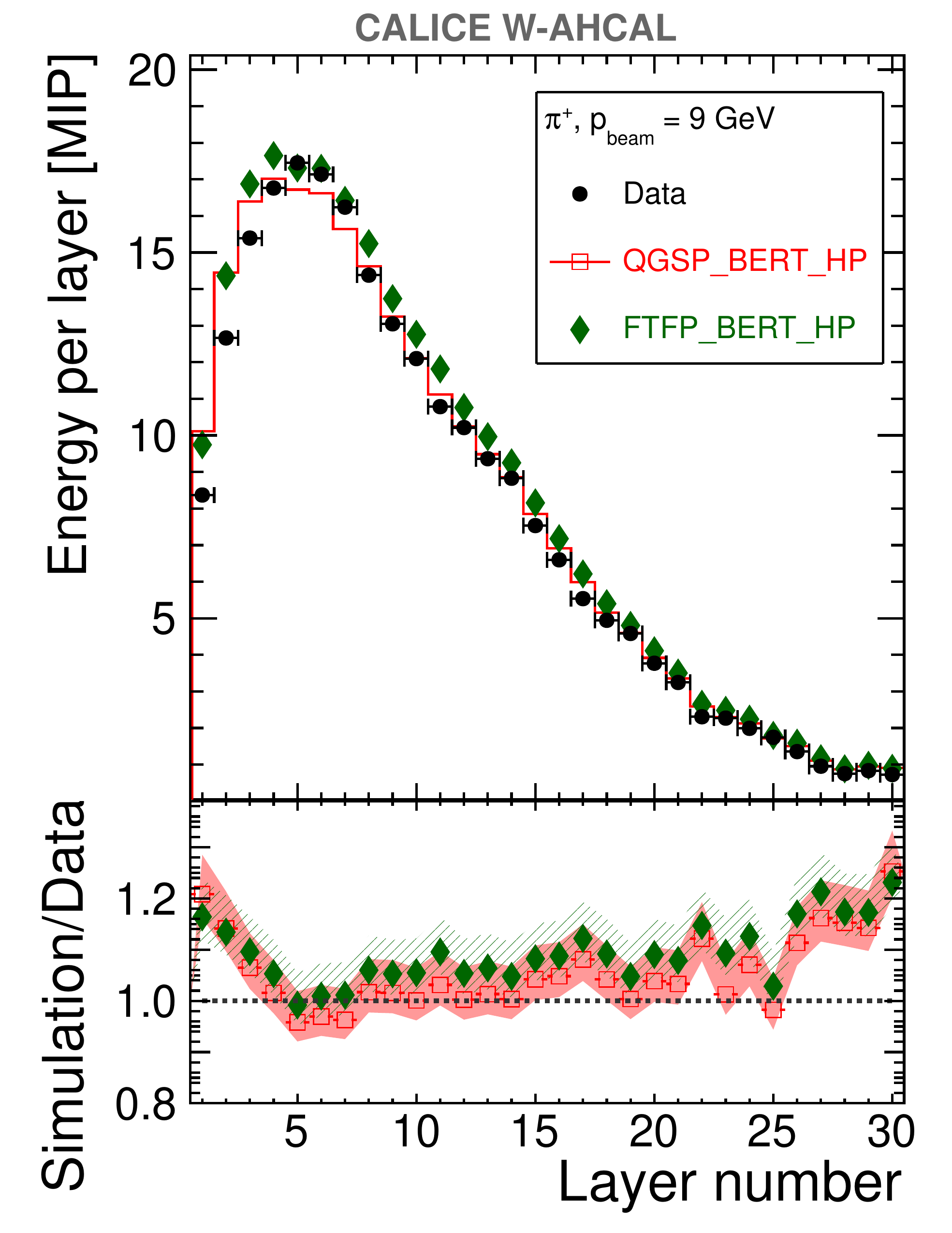}}
    \caption{(a) Longitudinal energy profile for pions at 10\,GeV measured in the Si-W ECAL~\cite{2015_SiWECAL}. Longitudinal energy profile for protons at 10\,GeV (b) and pions at 9\,GeV (c) measured in the W-AHCAL~\cite{2014_WAHCAL}. In all three figures several physics lists ({\sc Geant4}.9.6p01) are compared to the data.}
  }
\end{figure}

The depth of hadronic showers is evaluated by the longitudinal energy profile, shown for 10\,GeV pions in the Si-W ECAL in Fig.~\ref{subfig:longitudinalprofileECAL}, for 10\,GeV protons and 9\,GeV pions in the W-AHCAL in Fig.~\ref{subfig:longitudinalprofileHCALproton} and Fig.~\ref{subfig:longitudinalprofileHCALpion}.
For the Si-W ECAL the profile is evaluated from the start of the shower, which means the x-axis represents the shower depth.
The accuracy of the Monte Carlo simulations depends on the physics list used, as can be seen in Fig.~\ref{subfig:longitudinalprofileHCALproton} where QGSP\_BIC\_HP clearly performs less well than the other physics lists, but also on the version of {\sc Geant4}, as can be seen in Fig.~\ref{subfig:longitudinalprofileECAL}.
For the Si-W ECAL the newer version of {\sc Geant4} actually gives a worse description of the data. 
This has only been observed for silicon as active material.
Unexpected discrepancies like this are valuable input in improving the models in {\sc Geant4}.

Just like for the radial distributions, the longitudinal centre of gravity can be extracted from the profiles.
Simulations of this observable were found to reproduce the data very well, e.g. by FTFP\_BERT for the Fe-AHCAL, while QGSP\_BERT underestimates the data in the same prototype~\cite{2015_FeAHCAL}.
\vspace{-2mm}
\vspace{-2mm}

\begin{figure}
  {\centering
    \subfloat[\label{subfig:parametrised}]{\includegraphics[height=0.19\textheight]{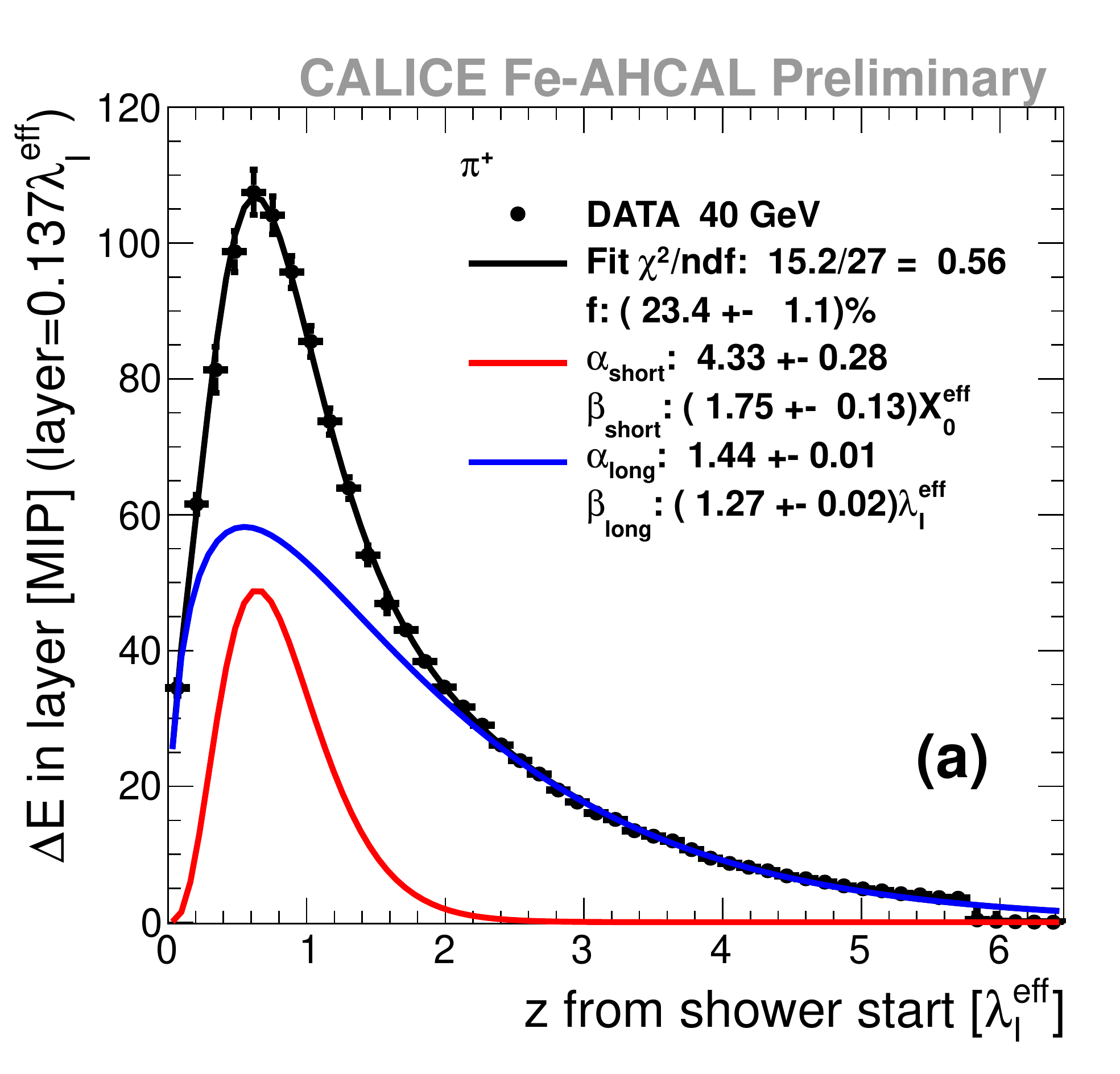}}
    \subfloat[\label{subfig:hovere}]{\includegraphics[height=0.19\textheight]{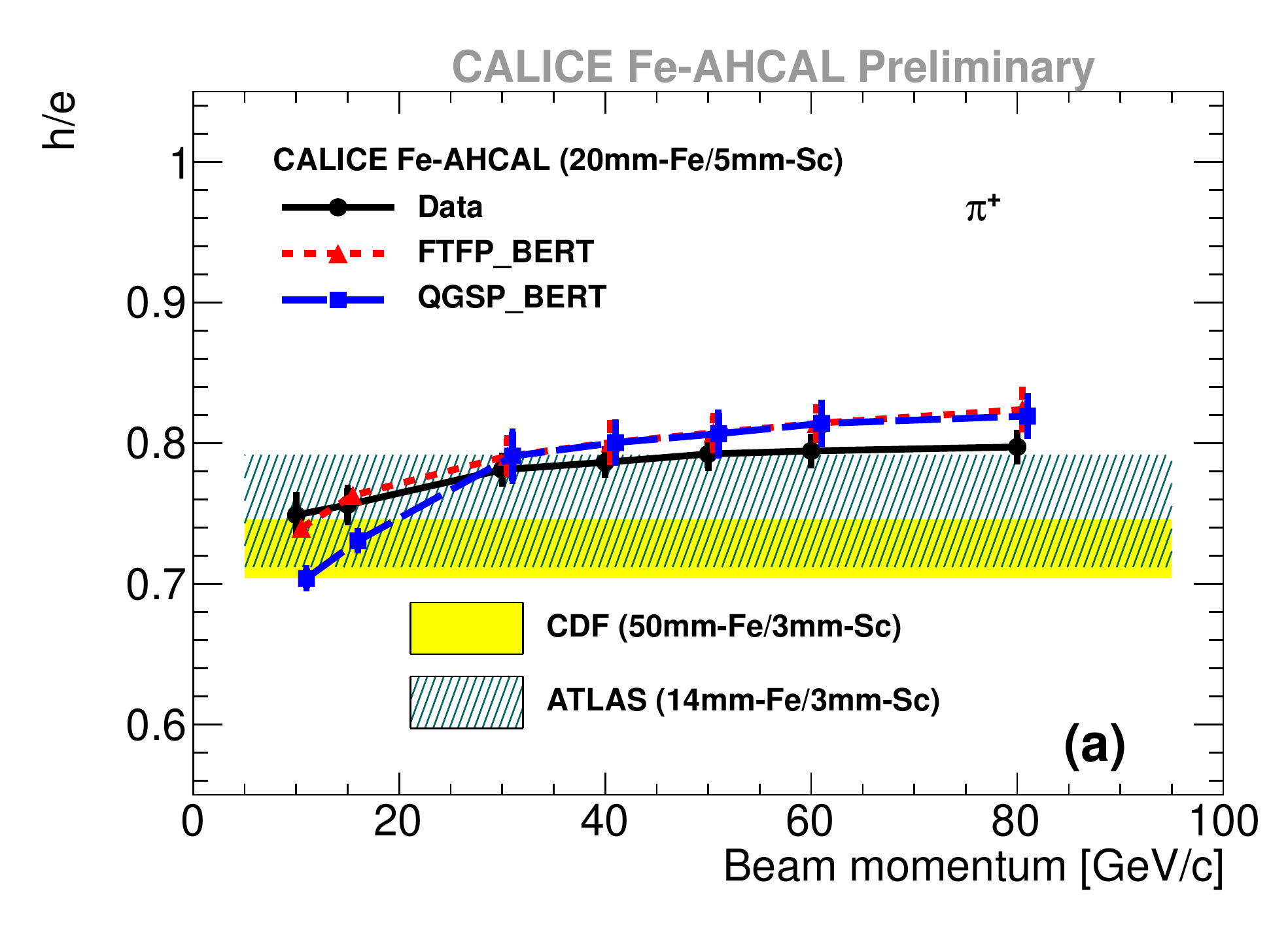}}
    \caption{(a) Parametrisation of the longitudinal energy profile for pions at 40\,GeV measured in the Fe-AHCAL~\cite{CAN048}. (b) The ratio of the hadronic to the electromagnetic response, h/e, for pions between 10\,GeV and 80\,GeV measured in the Fe-AHCAL~\cite{CAN051}. Several physics lists ({\sc Geant4}.9.6p01) are compared to the data. Additionally a comparison to results from other experiments is shown.}
  }
\end{figure}

The longitudinal shower profile, evaluated from the shower starting position, can be parametrised by a two-component function, as is illustrated in Fig.~\ref{subfig:parametrised} for pions in the Fe-AHCAL~\cite{CAN048}.
These two components show behaviour that can be associated to radiation length, and thus the electromagnetic part of the shower, and interaction length, and thus the pure hadronic part of the shower.
From these fit results the ratio of the hadronic to the electromagnetic response can be estimated.
It is shown in Fig.~\ref{subfig:hovere} for different energies and compared to results from other experiments (ATLAS and CDF)~\cite{CAN051}.
FTFP\_BERT describes the data best below 30\,GeV, at higher energies both the shown physics lists overestimate the data.
\vspace{-2mm}

\begin{figure}
  {\centering
    \subfloat[\label{subfig:delayed}]{\includegraphics[height=0.19\textheight]{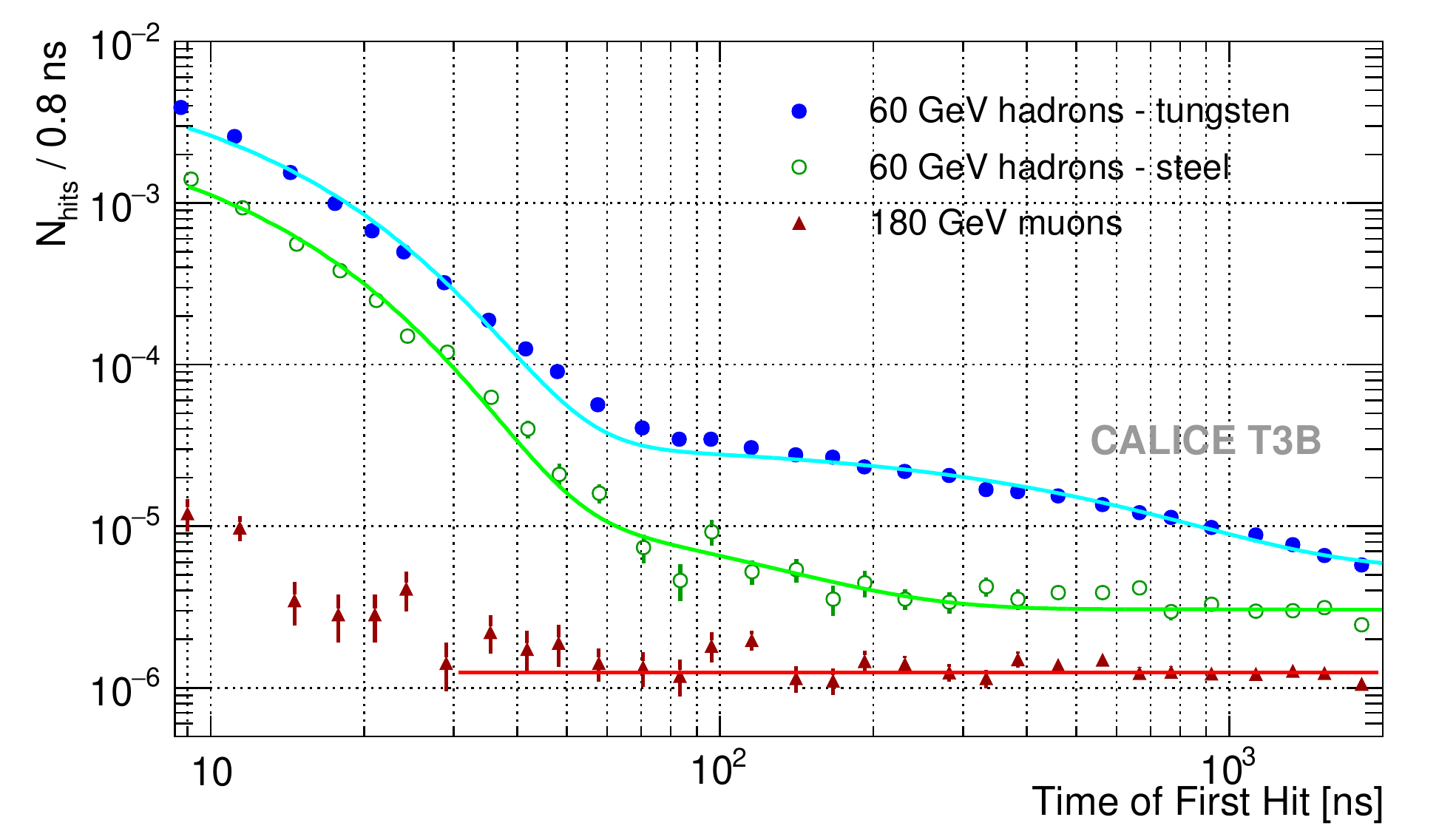}}
    \subfloat[\label{subfig:tofh}]{\includegraphics[height=0.19\textheight]{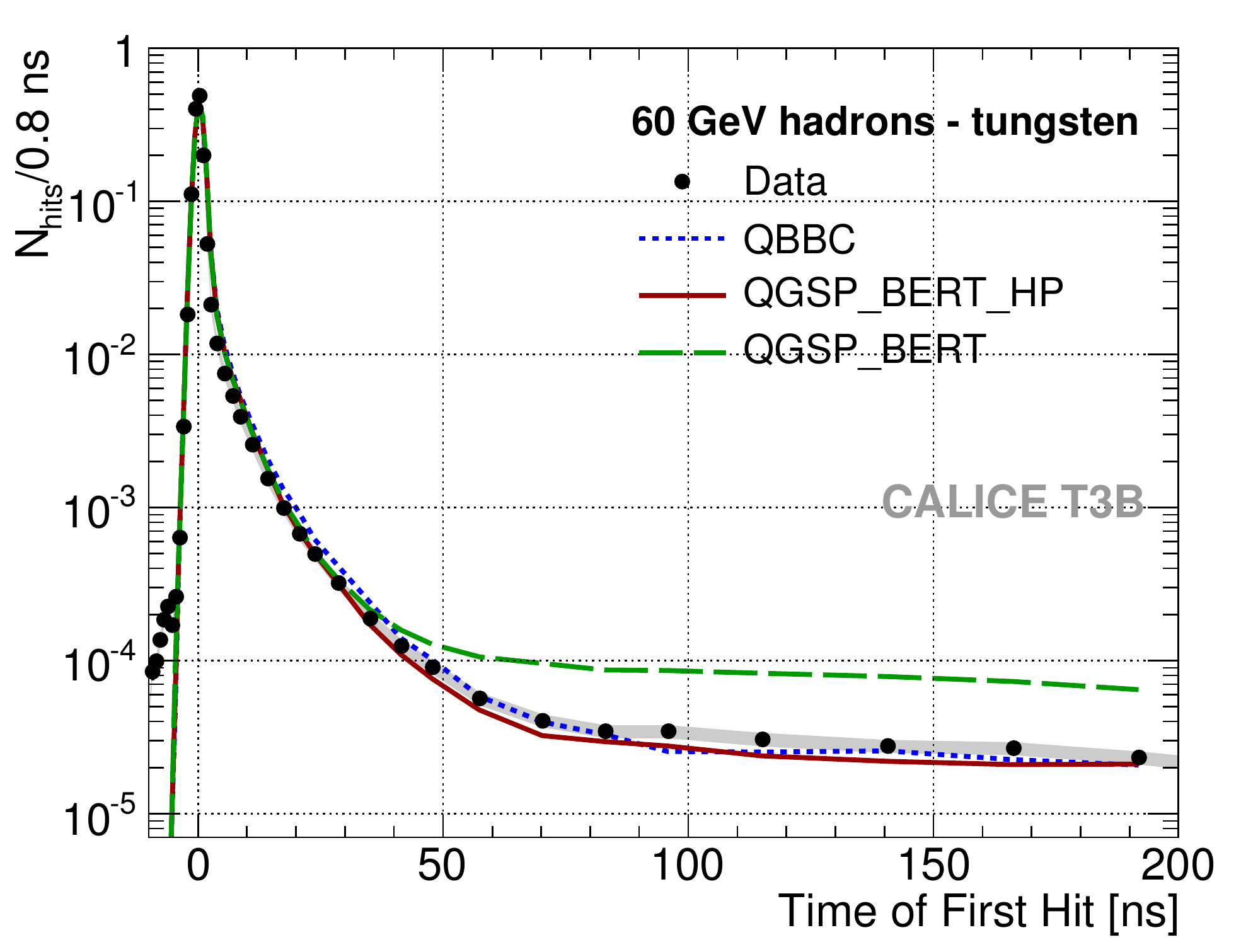}}
    \caption{(a) Delayed signals observed in T3B for steel and tungsten absorbers, compared to the instantaneous signal of muons, and (b) arrival time of hits in T3B for 60\,GeV hadrons with tungsten absorber~\cite{2014_T3B}. Several physics lists ({\sc Geant4}.9.4p03) are compared to the data.}
  }
\end{figure}

\vspace{-2mm}
The time structure within hadronic showers has been measured by the dedicated experiment T3B~\cite{2014_T3B}.
These studies are motivated by the fast event rates and pile up at the proposed CLIC accelerator, which makes timing cuts essential to reduce background.
The arrival time of signals in a small strip of scintillators was measured with nano second resolution, at the end of the W-AHCAL and SDHCAL, with tungsten and steel absorber, respectively.
Figure~\ref{subfig:delayed} shows the delayed signals, arriving later than 8 ns, that are due to slow neutron processes.
This delayed signal is more pronounced in tungsten than in steel, as expected because of the larger number of neutrons in its nucleus.
In the comparison of Monte Carlo to data, shown in Fig.~\ref{subfig:tofh} for 60 GeV hadrons in tungsten, it is clear that only the physics lists that incorporate a high precision treatment of low energy neutrons, QBBC and QGSP\_BERT\_HP, are able to reproduce the data accurately.
For steel the high precision modelling has much less influence.
These results have validated the fast detailed neutron treatment as implemented in QBBC, which is now standard in all physics lists starting from {\sc Geant4} version 10.0.

\begin{figure}
  {\centering
    \subfloat[\label{subfig:trackmultiplicity}]{\includegraphics[height=0.2\textheight]{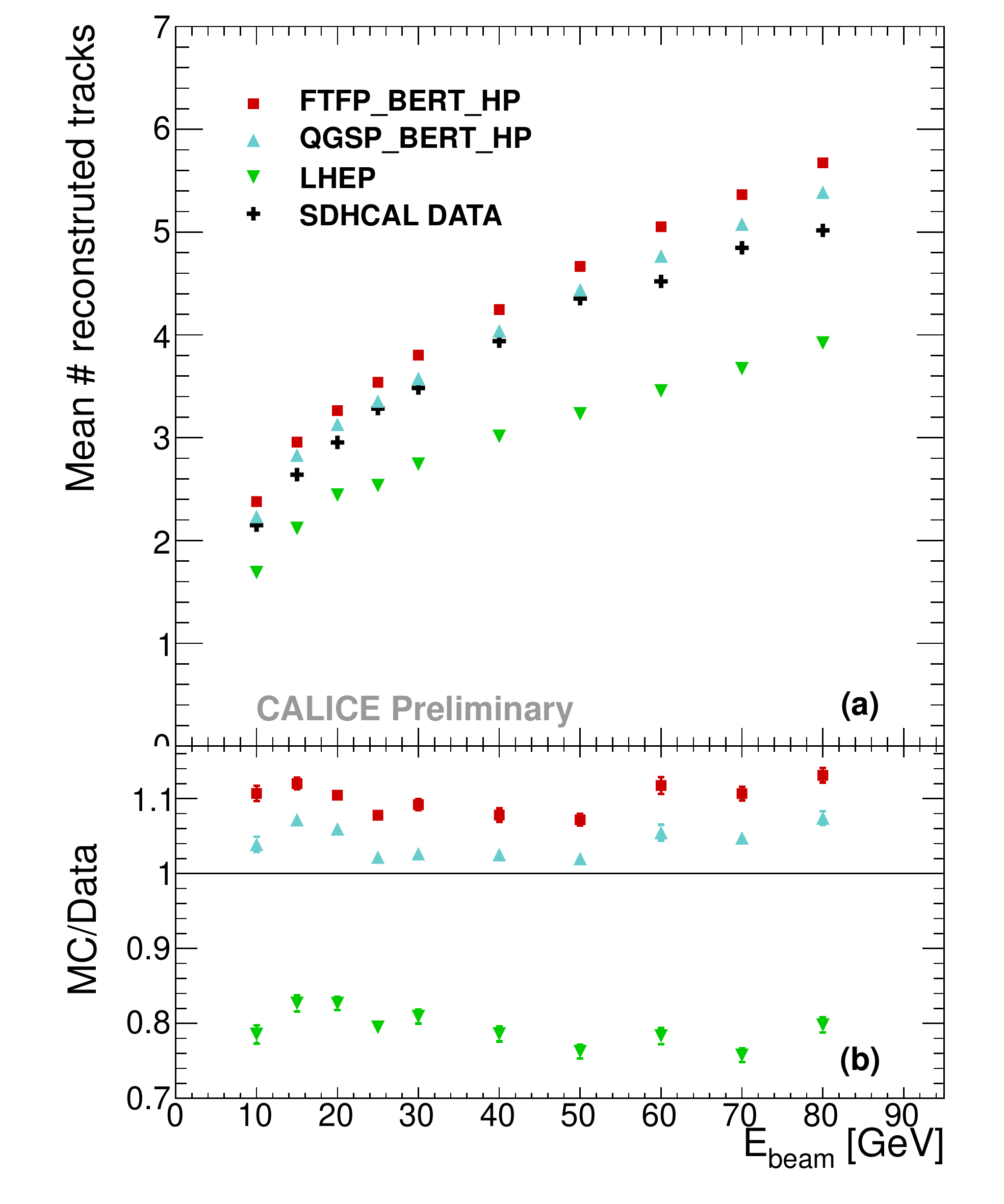}}
    \subfloat[\label{subfig:meantrackmult}]{\includegraphics[height=0.2\textheight]{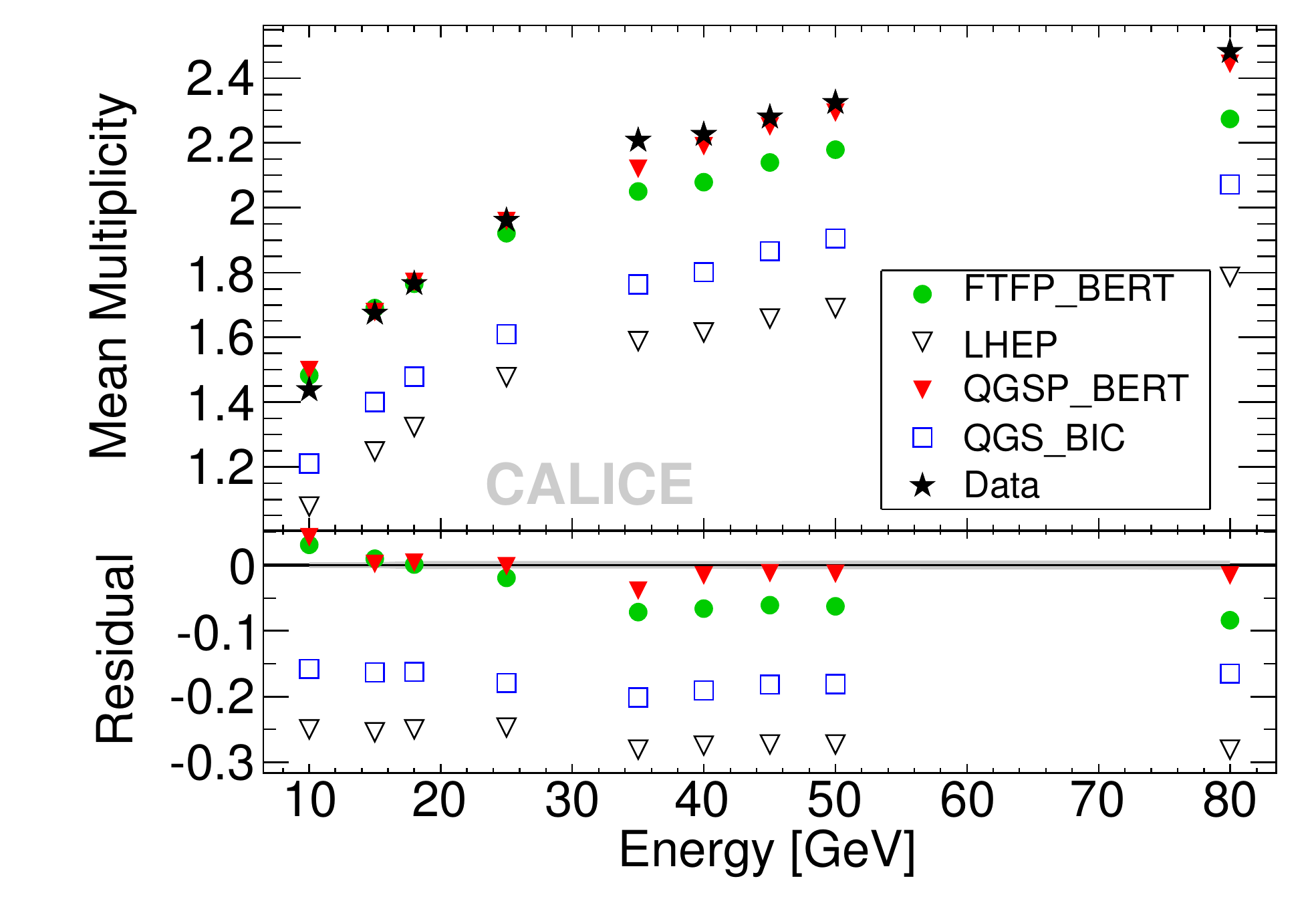}}
    \caption{The number of reconstructed tracks  in (a) the SDHCAL~\cite{CAN047} and (b) Fe-AHCAL~\cite{2013_FeAHCAL} .
    Several physics lists ((a) {\sc Geant4}.9.6p01, (b) {\sc Geant4}.9.4p02) are compared to the data.}
  }
  \label{fig:multiplicity}
\end{figure}

The detailed 3-dimentional data provided by highly granular calorimeters give access to the tree like structures of hadronic showers, where regions of high activity are connected by minimum ionising particle tracks.
These tracks can be reconstructed both in data and simulations.
A comparison of the number of tracks inside the shower has been made for e.g. pions in the SDHCAL~\cite{CAN047} and Fe-AHCAL~\cite{2013_FeAHCAL} as illustrated in Fig.~\ref{subfig:trackmultiplicity} and Fig.~\ref{subfig:meantrackmult}.
The physics list QGSP\_BERT reproduces the data best (within 5\%).
This level of detail in shower modelling aids in the 2-particle separation in Particle Flow Algorithms, which is an important constraint on the energy resolution that can be achieved by future experiments.
\vspace{-2mm}

\vspace{-2mm}

\section{Summary and Conclusion}
\vspace{-2mm}
In this conference contribution selected results of detailed measurements of hadronic showers recorded by the CALICE ECAL and HCAL prototypes are shown.
The comparison of these detailed 3-dimentional measurements to {\sc Geant4} models helps to improve the accuracy of hadronic shower simulations.
Current models reproduce the data within 5 to 20\%, depending on the physics list and version of {\sc Geant4}.
Generally, QGSP\_BERT and FTFP\_BERT perform best. 
The high precision (HP) treatment of neutrons is needed when simulating scintillator-tungsten prototypes.
The high level of detail that can be reproduced by modern simulations support the theory-driven approach adopted by the {\sc Geant4} developers.
The full potential of the CALICE data has not yet been exploited and increasingly detailed studies are undertaken.
This has the potential to lead to even better simulation models in future.
\vspace{-2mm}

\bibliographystyle{pos}
\bibliography{proceedings}

\begin{thebibliography}{10}
\expandafter\ifx\csname urlstyle\endcsname\relax
  \expandafter\ifx\csname doi\endcsname\relax
  \def\doi#1{doi:\discretionary{}{}{}#1}\fi \else
  \expandafter\ifx\csname doi\endcsname\relax
  \def\doi{doi:\discretionary{}{}{}\begingroup \urlstyle{rm}\Url}\fi \fi

\bibitem{PoS_Vladik}
{V. Balagura for the CALICE collaboration}, {\em {D}evelopment of technologies
  for highly granular calorimeters and their performance in beam tests\/}, {\em
  {\em in proceedings of} {T}he {E}uropean {P}hysical {S}ociety on {H}igh
  {E}nergy {P}hysics\/} {\bf \pos{PoS(EPS-HEP2015)245}}.

\bibitem{2009_Thomson}
M.~A. Thomson, {\em Particle {F}low {C}alorimetry and the {P}andora{PFA}
  {A}lgorithm\/}, {\em Nucl. Instrum. Meth. A\/} {\bf 611} 25,
  [arXiv:0907.3577].

\bibitem{2012_Marshall}
J.~S. Marshall, A.~M{\"u}nnich, and M.~A. Thomson, {\em {P}erformance of
  particle flow calorimetry at {CLIC}\/}, {\em Nucl. Instrum. Meth. A\/} {\bf
  700} 153, [arXiv:1209.4039].

\bibitem{2001_Brient}
J.~C. Brient and H.~Videau, {\em {T}he calorimetry at the future $e^{+}e^{-}$
  linear collider\/}, {\em {\em in proceedings of} The APS / DPF / DPB Summer
  Study on the Future of Particle Physics (Snowmass 2001)\/} [hep-ex/0202004].

\bibitem{2013_TDR4}
T.~Behnke {\em et~al.\/}, {\em The {I}nternational {L}inear {C}ollider
  {T}echnical {D}esign {R}eport - {V}olume 4: {D}etectors\/}, {\em
  ILC-REPORT-2013-040\/} [arXiv:1306.6329].

\bibitem{2012_CLIC}
L.~Linssen {\em et~al.\/}, {\em {P}hysics and {D}etectors at {CLIC}: {CLIC}
  {C}onceptual {D}esign {R}eport\/}, {\em CERN-2012-003\/} [arXiv:1202.5940].

\bibitem{2003_Geant4}
{The {\sc geant4} collaboration}, {\em {\sc geant4} -- a simulation toolkit\/},
  {\em Nucl. Instrum. Meth. A\/} {\bf 506} 250,
  http://geant4.web.cern.ch/geant4.

\bibitem{2010_Apostolakis}
J.~Apostolakis {\em et~al.\/}, {\em Validation of {\sc geant4} hadronic models
  using {CALICE} data\/}, {\em EUDET-MEMO-2010-15\/}
  {h}ttp://www.eudet.org/e26/e28/e86887/e109012/EUDET-Memo-2010-15.pdf.

\bibitem{2011_Dotti}
A.~Dotti {\em et~al.\/}, {\em Description of {H}adron-induced {S}howers in
  {C}alorimeters using the {\sc geant4} {S}imulation {T}oolkit\/}, {\em {\em in
  proceedings of} The IEEE NSS MIC 2011 Conference\/}
  {h}ttp://geant4.web.cern.ch/geant4/results/papers/hadronic-showers-IEEE11.pdf.

\bibitem{2015_SiWECAL}
{The CALICE collaboration}, B.~Bilki, {\em et~al.\/}, {\em {T}esting {H}adronic
  {I}nteraction {M}odels using a {H}ighly {G}ranular {S}ilicon-{T}ungsten
  {C}alorimeter\/}, {\em Nucl. Instrum. Meth. A\/} {\bf 794} 240,
  [arXiv:1411.7215].

\bibitem{2015_FeAHCAL}
{The CALICE collaboration}, B.~Bilki, {\em et~al.\/}, {\em Pion and proton
  showers in the {CALICE} scintillator-steel analogue hadron calorimeter\/},
  {\em J. Instrum.\/} {\bf 10}~(P04014), [arXiv:1412.2653].

\bibitem{2013_FeAHCAL}
{The CALICE collaboration}, C.~Adloff, {\em et~al.\/}, {\em {V}alidation of
  {\sc geant4} {M}onte {C}arlo {M}odels with a {H}ighly {G}ranular
  {S}cintillator-{S}teel {H}adron {C}alorimeter\/}, {\em J. Instrum.\/} {\bf
  8}~(P07005), [arXiv:1306.3037].

\bibitem{CAN044}
{The CALICE collaboration}, {\em Shower development of particles with momenta
  from 10 to 100 {GeV} in the {CALICE} {S}cintillator-{T}ungsten {HCAL}\/},
  {\em CALICE Analysis Note\/} {\bf CAN-044},
  https://twiki.cern.ch/twiki/pub/CALICE/CaliceAnalysisNotes/CAN-044.pdf.

\bibitem{2014_WAHCAL}
{The CALICE collaboration}, C.~Adloff, {\em et~al.\/}, {\em {S}hower
  development of particles with momenta from 1 to 10 {GeV} in the {CALICE}
  {S}cintillator-{T}ungsten {HCAL}\/}, {\em J. Instrum.\/} {\bf 9}~(P01004),
  [arXiv:1311.3505].

\bibitem{CAN048}
{The CALICE collaboration}, {\em Parametrisation of hadron shower profiles in
  the {CALICE} {S}c-{F}e {AHCAL}\/}, {\em CALICE Analysis Note\/} {\bf
  CAN-048},
  https://twiki.cern.ch/twiki/pub/CALICE/CaliceAnalysisNotes/CAN-048.pdf.

\bibitem{CAN051}
{The CALICE collaboration}, {\em Extraction of h/e and calorimeter response
  from fits to the longitudinal shower profiles in the {CALICE} {S}c-{F}e
  {AHCAL}\/}, {\em CALICE Analysis Note\/} {\bf CAN-051},
  https://twiki.cern.ch/twiki/pub/CALICE/CaliceAnalysisNotes/CAN-051.pdf.

\bibitem{2014_T3B}
{The CALICE collaboration}, C.~Adloff, {\em et~al.\/}, {\em The {T}ime
  {S}tructure of {H}adronic {S}howers in highly granular {C}alorimeters with
  {T}ungsten and {S}teel {A}bsorbers\/}, {\em J. Instrum.\/} {\bf 9}~(P07022),
  [arXiv:1404.6454].

\bibitem{CAN047}
{The CALICE collaboration}, {\em Tracking within {H}adronic {S}howers in the
  {SDHCAL} prototype using {H}ough {T}ransform {T}echnique\/}, {\em CALICE
  Analysis Note\/} {\bf CAN-047},
  https://twiki.cern.ch/twiki/pub/CALICE/CaliceAnalysisNotes/CAN-047.pdf.

\end{thebibliography}

\end{document}